\documentclass[useAMS,usenatbib,fleqn]{mn2e}
\usepackage{amsmath}
\usepackage{array}
\usepackage{booktabs}
\usepackage{xifthen}
\usepackage{graphicx}
\usepackage{color}
\usepackage{xspace}
\usepackage{url}
\usepackage{amssymb}
\usepackage{mathtools} 
\usepackage{enumitem}
\usepackage{natbib}

\def \aap{A\&A}

\def \apjl{ApJ}
\def \apjs{ApJS}

\newcommand{\reb}{{\sc \tt REBOUND}\xspace}

\newcommand{\whfast}{{\sc \tt WHFast}\xspace}
\newcommand{\ias}{{\sc \tt IAS15}\xspace}

\newcommand{\bo}[1][]{%
    \ifthenelse{\equal{#1}{}}{\mathcal{O}}{\mathcal{O}\left(#1\right)}%
}

\def\mc#1{{\mathcal{#1}}}

\title[High order symplectic integrators in REBOUND]{High order symplectic integrators for planetary dynamics and their implementation in REBOUND}
\date{Submitted: 25 July 2019; Accepted: 4 September 2019}

\author[Rein et al.]{Hanno Rein$^{1,2,3}$, 
    Daniel Tamayo$^{4}$\thanks{NHFP Sagan Fellow},
    Garett Brown$^{1,3}$\\
$^1$ Department of Physical and Environmental Sciences, University of Toronto at Scarborough, Toronto, Ontario M1C 1A4, Canada\\
$^2$ Department of Astronomy and Astrophysics, University of Toronto, Toronto, Ontario, M5S 3H4, Canada\\
$^3$ Department of Physics, University of Toronto, Toronto, Ontario, M5S 3H4, Canada,\\
$^4$ {Department of Astrophysical Sciences, Princeton University, Princeton, New Jersey 08544, United States}\\
}

\vspace{0.5\baselineskip}

\begin{document}
\maketitle

\begin{abstract}
    Direct N-body simulations and symplectic integrators are effective tools to study the long-term evolution of planetary systems.  The Wisdom-Holman (WH) integrator in particular has been used extensively in planetary dynamics as it allows for large timesteps at good accuracy.  One can extend the WH method to achieve even higher accuracy using several different approaches.  In this paper we survey integrators developed by Wisdom et al. (1996), Laskar \& Robutel (2001), and Blanes et al. (2013).  Since some of these methods are harder to implement and not as readily available to astronomers compared to the standard WH method, they are not used as often.  This is somewhat unfortunate given that in typical simulations it is possible to improve the accuracy by up to six orders of magnitude (!) compared to the standard WH method without the need for any additional force evaluations.  To change this, we implement a variety of high order symplectic methods in the freely available N-body integrator REBOUND.  In this paper we catalogue these methods, discuss their differences, describe their error scalings, and benchmark their speed using our implementations.
\end{abstract}


%%%%%%%%%%%%%%%%%%%%%%%%%%%%%%%%%%%%%%%%%%%%%%%%%
%%%%%%%%%%%%%%%%%%%%%%%%%%%%%%%%%%%%%%%%%%%%%%%%%
\section{Introduction}
\label{sec:intro}
Astronomers have predicted the location of planets in the night sky since ancient times.
However, it has only recently become possible to calculate the orbital evolution of planetary systems accurately over very long time scales with the help of fast computers \citep{LaskarGastineau2009}.
Various different numerical algorithms, so called integrators, have been used for that purpose.
Because of the large separation of timescales involved, from a day to billions of years, specialized integrators are needed to predict the orbital evolution of planetary systems over their entire lifetime which can correspond to up to $10^{12}$ orbits.

There are many options when it comes to choosing an integrator.
Some generic integrators achieve a good accuracy by being very high order.
Examples include the Bulirsch-Stoer \citep{StoerBulirsch02} or the Gau\ss-Radau based \ias integrator \citep{ReinSpiegel2015}.
A different approach is that of \cite{WisdomHolman1991}, which makes use of the fact that we are interested in the evolution of a dynamical system rather than simply the solution of a generic differential equation, and that the gravitational interactions between planets can be considered perturbations to otherwise Keplerian orbits.
Their integrator, which we refer to a the classical Wisdom-Holman integrator (WH) is therefore particularly well suited for planetary systems where orbits remain well separated.
This is the case for many planetary systems including the Solar System.
Different implementations of the WH integrator and various extensions of it (for example allowing close encounters) are freely available and have been used extensively by the astrophysics community \citep{Chambers1997,Duncan1998,ReinTamayo2015,Rein2019}. 

The classical WH integrator is a second order method. 
If one requires greater accuracy one has two options: either reducing the timestep or choosing a different higher order method (which ideally also makes use of the fact that planet-planet interactions are small).
Many simulations require an astronomically large number of timesteps to begin with\footnote{Whereas in the Solar System the shortest orbital period is 88~days, there are many extrasolar planets with extremely short periods. Integrating a planet on a one day orbit for 10~Gyrs requires roughly~$10^{14}$~timesteps. } and it is not uncommon for simulations to run for a month or even a year in some cases \citep{La2010}.
Since reducing the timestep, and thus increasing the number of timesteps, leads to a greater computational cost there might be a significant advantage if one can use a higher order method which offers better accuracy at a fixed timestep. 

The higher order integrators that we are looking at in this paper all assume that planetary systems do not have close encounters and planetary orbits remain well separated.
There are once again different approaches for obtaining such a high order integrator. 
Each comes with many subtle choices that one can make along the way.
In this paper we focus on the advances driven by two groups, one led by Jack Wisdom \citep{Wisdom1996, Wisdom2006, Wisdom2018} and the other by Jacques Laskar \citep{LaskarRobutel2001,Blanes2013}.
Since their implementations are not easily available for the community, we have implemented their integrators within the \reb package, together with some new variants.
We review the different approaches that these groups take in Sect.~\ref{sec:review} and compare their integrators' formal properties.
We present our own implementations in Sect.~\ref{sec:implementation}.
We then verify the accuracy of our implementations and measure their speed in numerical tests that we present in Sec.~\ref{sec:tests}. 
We hope that making these integrators easily available within \reb will allow more people who are interested in planetary dynamics to use and build upon them.

%%%%%%%%%%%%%%%%%%%%%%%%%%%%%%%%%%%%%%%%%%%%%%%%%
%%%%%%%%%%%%%%%%%%%%%%%%%%%%%%%%%%%%%%%%%%%%%%%%%
\section{Symplectic Integrators}
\label{sec:review}
%%%%%%%%%%%%%%%%%%%%%%%%%%%%%%%%%%%%%%%%%%%%%%%%%
\subsection{Splitting methods}
\label{sec:splitting}
All integrators described in this paper are symplectic splitting schemes.
In this section, we provide a short introduction to such methods.
This is by no means a comprehensive discussion of this vast subject area.
Our goal is not to rederive these methods but to introduce the notations that we will use later.
We opt for a formal but nevertheless hopefully easy to understand approach for those with little background in this area. 
For different approaches and perspectives we refer the reader to papers and books by \cite{LaskarRobutel2001,Wisdom2018,Hairer2006} as well as references therein.

The differential equations of the $N$-body problem have special properties because they can be derived from a Hamiltonian $H$ together with Hamilton's equations.
From a mathematical point of view, such differential equations have more structure than an arbitrary one.
Many of these mathematical properties have direct physical consequences, like the conservation of energy and phase space volume.
The integrators that we describe in this paper preserve some of these structures by construction, and one can argue that they are therefore preferable.
We note, however, that even though we focus on conservative systems here, many of the splitting methods that we describe can also be applied to non-Hamiltonian systems \citep{Tamayo2019}.

Let us first define the Poisson bracket $\{g,h\}$ of two functions $g$ and $h$  of the canonical coordinates $(q_i, p_i)$ as
\begin{eqnarray}
    \{g,h\} =  \sum_i\left(  \frac{\partial g}{\partial q_i } \frac{\partial h}{\partial p_i }-  \frac{\partial g}{\partial p_i } \frac{\partial h}{\partial q_i }\right).
\end{eqnarray}
The functions $g$ and $h$ are arbitrary, they can for example be a single coordinate, the energy, or the angular momentum.
We can use the Poisson bracket to write down Hamilton's equations as
\begin{eqnarray}
    \dot q_i = \frac{\partial H}{\partial p_i} = \{q_i,H\}\quad\quad\text{and}\quad\quad
    \dot p_i = -\frac{\partial H}{\partial q_i} = \{p_i,H\}.
\end{eqnarray}
Thus, through the chain rule, the time derivative of any function $g$ that depends only on $p$ and $q$ can be written succinctly as
\begin{eqnarray}
    \dot g = \{g,H\}.  \label {gH}
\end{eqnarray}
Assume that the planetary system we are interested in has $3N$ coordinates and $3N$ momenta.
If we define a function $y(t)=(q_1(t),\ldots,q_{3N}(t), p_1(t),\ldots,p_{3N}(t))$, then solving the $N$-body problem corresponds to solving the $3N$ coupled differential equations $\dot y = \{y,H\}$.
Let us further define the Lie derivative with respect to the Hamiltonian $H$ as 
\begin{eqnarray}
    \mc L_H y = \{y,H\}, 
\end{eqnarray}
or, just writing down the operator without the operand, 
\begin{eqnarray}
    \mc L_H  = \{\cdot, H\}.
\end{eqnarray}
Comparing with Eq.\:\ref{gH}, we see that the Lie derivative with respect to the Hamiltonian $H$ is an operator that yields the time derivative of any function of the phase space coordinates that it acts on.
Finally, let us define the formal solution operator $\varphi^{[H]}_t(y_0)$.
This operator returns the solution to the differential equation $\dot y = \{y,H\}$ at time $t$ with initial conditions $y_0$ given at $t=0$.

In general we do of course not know how to write down this abstract solution operator.
However, with the notation that we have introduced, we can now at least write down a formal expression.
In terms of the Lie derivative defined above and an exponential defined in the usual way as an infinite series, a formal expression of the solution operator is
\begin{eqnarray}
    {\varphi}^{[H]}_{t}(y_0) &=& \exp\left(t \mc L_{H} \right) \text{Id}(y_0)\label{eq:expsol}\\
    &=& \nonumber
     \left(\text{Id} + t \mc L_H  \text{Id}+\frac12 \mc L_H \mc L_H \text{Id} +\ldots \right) (y_0). 
\end{eqnarray}
Note that the exponential in Eq.~\ref{eq:expsol} is simply a formal way to write down an operator.
It is acting on the function to its right, in the above case this is the identity $\text{Id}$.
To see that this construct indeed provides a solution to the differential equation $\dot y = \{y,H\}$ with initial conditions $y_0$, note that the series expansion of the exponential results in a series of differential operators acting on $\text{Id}(y_0)$. 
This series is simply the Taylor expansion of the solution ${\varphi}$ around $t=0$. 

Since it is not possible to write down an explicit solution for the $N$-body problem, we turn to operator splitting methods.
One way to view an operator splitting method is to take a differential equation of the form
\begin{eqnarray}
    \dot y = \mc L_H (y) \label{eq:ode}
\end{eqnarray}
and rewrite it as 
\begin{eqnarray}
    \dot y = \left(\mc L_A +  \mc L_B\right)(y)  =  \mc L_A(y)+   \mc L_B(y)  \label{eq:deq}.
\end{eqnarray}
Note that due to the linearity of the Poisson bracket, we have $\mc L_{A+B} = \mc L_A + \mc L_B$.
This implies that for Hamiltonian systems we can \emph{split} the equations of motion by \emph{splitting} the Hamiltonian $H$ into two parts, i.e. $H=A+B$.
Depending on the splitting, we get two new sets of differential equations which we can study independently:
\begin{eqnarray}
    \dot y = \mc L_A (y) \quad \text{and} \quad \dot y =  \mc L_B (y) \label{eq:spliteq}.
\end{eqnarray}
In practice we will choose a splitting which allows us to easily solve these new differential equations independently. 
In other words, we want to be able to explicitly write down the solution operators ${\varphi}^{[A]}_{t}(y_0)$ and ${\varphi}^{[B]}_{t}(y_0)$. 

Using the notation from above, we have
\begin{eqnarray}
    \exp\left(t \mc L_{H} \right)
    = 
    \exp\left(t (\mc L_{A}+\mc L_B) \right),
\end{eqnarray}
but because these exponentials are operators, not scalars, we unfortunately can't use standard rules for exponentials.
Specifically, we typically have
\begin{eqnarray}
    \exp\left(t (\mc L_{A}+\mc L_B) \right)
    \neq 
    \exp\left(t \mc L_{A} \right)
    \exp\left(t \mc L_{B} \right).
\end{eqnarray}
For illustration purposes, imagine the above equation were true. 
We could then solve the differential equation in Eq.~\ref{eq:ode} by first applying the solution operator ${\varphi}^{[B]}$ to our initial conditions for a time $t$ (remember, operators act to the right), then take the result and apply the solution operator ${\varphi}^{[A]}$ on it, again for time $t$.
The result of the last operation would then correspond to the solution ${\varphi}^{[H]}$ at time $t$. 
But because the equation above is not true, we need to rely on the Baker-Campbell-Hausdorff (BCH) identity\footnote{There is a small subtlety which originates from a permutation of the exponentials known as the Vertauschungssatz \citep{Grobner1967}. This has occasionally led to to a sign error in the literature. The fact that there is no sign ambiguity at even orders did not help to clear up the confusion. See Lemma 5.1 in chapter III of \cite{Hairer2006}.}, to get a similar expression
\begin{eqnarray}
    &&  \exp\left(t \mc L_{A} \right)
    \exp\left(t \mc L_{B} \right) \nonumber \\
    && =
    \exp\left(t \mc L_{A} + t \mc L_{B} + \frac12 t^2 \left[\mc L_A,\mc L_B \right] + \bo[t^3]\right), \label{eq:bch}
\end{eqnarray}
where $[\mc L_A, \mc L_B] =\mc L_A \mc L_B - \mc L_B \mc L_A $ is a commutator. 
So if we apply the splitting procedure described above (corresponding to the left hand side of Eq.~\ref{eq:bch}), we do not solve the differential equation $\dot y = \mc L_H (y)$, but some other differential equation (given by the argument in the exponential on the right hand side of Eq.~\ref{eq:bch}).
Formally, we can write down the Hamiltonian corresponding to the differential equation that we are actually solving as 
\begin{eqnarray}
    H' = A+B+\frac12 t \{A,B\} + \bo[t^2]
\end{eqnarray}
where we have used 
\begin{eqnarray}
    \left[ \mc L_{A} , \mc L_{B} \right] = \mc L_{\{A,B\}},
\end{eqnarray}
which follows from the Jacobi identity. 
Although we do not end up solving the Hamiltonian system we were trying to solve, the fact that we can write down a different Hamiltonian system that we seem to solve exactly is often quoted as why symplectic integrators have superior properties\footnote{
As \cite{Wisdom2018} explains, this argument is flawed because it ignores that fact that the series above might not converge everywhere in phase space.
The system we end up actually solving might look nothing like the system we want to solve.
The interpretation preferred by \cite{Wisdom2018} avoids some of these problems.
However, for the discussion in this paper, this is not a concern.}.
We can study the properties of this `nearby' Hamiltonian to see how far our approximation is from the true solution. 
As we can see from the equation above, for small $t$, the Hamiltonian is close to the original Hamiltonian and only differs by  $\frac12 t \{A,B\} +\bo[t^2]$.
The smaller a step we make, the closer we stay to the true Hamiltonian.
We indicate that we take a small but finite step by replacing $t$ with $dt$.
To reach a time~$t$ we simply repeat the procedure $n$ times such that $t=dt\,n$.

%%%%%%%%%%%%%%%%%%%%%%%%%%%%%%%%%%%%%%%%%%%%%%%%%
\subsection{Wisdom-Holman integrator}
\cite{WisdomHolman1991} split the $N$-body Hamiltonian~$H$ of a planetary system into two parts~$A$ and~$B$ such that~$A$ contains all terms describing the planets' Keplerian motion and~$B$ contains the planet-planet interactions. 
Jacobi coordinates are particularly well suited for this.
This coordinate choice has a further advantage, it keeps the momentum terms in the Hamiltonian in the form $\sum_i p_i^2/(2m_i)$.
For an analysis of other splittings, see \cite{HernandezDehnen2016} and \cite{ReinTamayo2019}.

Note that the Hamiltonian $B$ only contains a position-dependent potential. 
We can easily calculate the corresponding equations of motion, $\dot y = \mc L_B y$, and solve them exactly.
The solution  operator ${\varphi}^{[B]}$ is simply a kick, i.e. a change of the particles' momenta while keeping their positions fixed.
Although it is not quite as straightforward, the solution ${\varphi}^{[A]}$ to the equations of motion $\dot y= \mc L_A y$ can also be computed relatively easily if one can efficiently solve Kepler's equation \citep[the solution is only required to within finite machine precision, see e.g.][]{ReinTamayo2015}.

The motion of well separated planets is described almost completely by only the Keplerian motion,~$A$, and we can consider the interactions due to other planets,~$B$, a perturbation. 
We can quantify this with a small number $\epsilon$ and keep track of it by formally replacing $B$ in the Hamiltonian with $\epsilon B$ such that the splitting is of the form $H=A+\epsilon B$. 

The splitting scheme described in the previous section is a first order scheme, the WH integrator uses a second order one.
To advance the solution by one timestep $dt$, it first applies the solution operator ${\varphi}_{dt/2}^{[A]}$ for half a timestep, then applies ${\varphi}_{dt}^{[B]}$ for a full timestep, followed by another operation of ${\varphi}_{dt/2}^{[A]}$.
Let us formally write one timestep of this scheme as
\begin{eqnarray}
    \text{WH} = e^{\frac12 A}e^{ B}e^{\frac12 A}.
\end{eqnarray}
As before, we can apply the BCH formula (twice in this case) to calculate the nearby Hamiltonian, i.e. the Hamiltonian of the system the integrator is actually solving (sometimes this is also referred to as the `shadow' Hamiltonian), to be
\begin{eqnarray}
    &&  H' = 
     A +  \epsilon B  \nonumber
    - \epsilon \frac{dt^2}{24} \left\{A,\left\{A, B \right\} \right\}
    - \epsilon^2 \frac{dt^2}{12} \left\{B,\left\{A, B \right\} \right\}\\
    &&+ \bo(\epsilon dt^4) 
    + \bo(\epsilon^2 dt^4)
    + \ldots
\end{eqnarray}
Note that the first error terms appear at second order in~$dt$.
There are two terms, one is $\bo[\epsilon dt^2]$, the other is $\bo[\epsilon^2 dt^2]$.
For small $\epsilon$, the first term dominates.
Compared to other methods, for example the standard leap-frog integrator\footnote{The leap-frog integrator is also a second order symplectic splitting method, but splits the Hamiltonian into a potential and kinetic term. Both terms are both equally important to describe the motion of a planet, thus $\epsilon\sim1$.}, the thing to note here is the factor of $\epsilon$. 
For the Solar System, $\epsilon\sim10^{-3}$, making the WH integrator about 1000~times more accurate than the leap-frog integrator whose error terms have the same structure as those above, but do not include the factors of $\epsilon$.

We refer to this integrator as the classical Wisdom-Holman method, WH.

%%%%%%%%%%%%%%%%%%%%%%%%%%%%%%%%%%%%%%%%%%%%%%%%%
\subsection{Wisdom-Holman-Touma integrator family}
\label{sec:WHT}

%%%%%%%%%%%%%%%%%%%%%%%%%%%%%%%%%%%%%%%%%%%%%%%%%
\subsubsection{Symplectic correctors}
As we've seen above, the classical WH method has two error terms at second order in time, $\bo[\epsilon dt^2]$ and $\bo[\epsilon^2 dt^2]$.
To obtain a higher accuracy method, we would like to get rid of the leading terms.
Instead of simply using a higher order splitting scheme \citep{Yoshida1993}, \cite{Wisdom1996} describe a different approach where in their derivation they make use of the fact that the system we are integrating is a Hamiltonian system.

\cite{Wisdom1996} apply a corrector $e^C$ step before and after each step to remove the leading term, $\bo[\epsilon dt^2]$. Specifically the method is
\begin{eqnarray}
    \text{WHC} = e^{ C} e^{\frac12 A}e^{  B}e^{\frac12 A} e^{-C}.
\end{eqnarray}
Because $e^{- C}$ is the inverse of $e^{ C}$, these correctors only need to be applied at the beginning of the integration and at the end, or whenever an output is generated.
But they do not need to be applied in-between timesteps since they cancel out.
For long integrations this method is therefore effectively as fast as the classical WH method.

Deriving the $e^{C}$ operator would go beyond the scope of this paper and we refer the reader to \cite{Wisdom1996} and \cite{Wisdom2018}.
To summarize, the leading-order error term (at order $\epsilon$) in the classical WH method can be interpreted as arising from a mismatch in initial conditions between the real problem and the modified problem that the splitting scheme is solving. 
Applying the symplectic correctors corresponds to a canonical transformation to and from integrator coordinates which removes the $\mathcal{O}(\epsilon)$ term\footnote{Given that the difference between real and integrator coordinates also depends on the timestep (and vanishes in the limit $dt \rightarrow 0$), \cite{SahaTremaine1992} show that a `warm-up' procedure that slowly increases the timestep from 0 to the desired finite value $dt$ achieves essentially the same result as an explicit symplectic corrector to integrator variables. 
We do not discuss this method further because it has almost identical properties to the ones we describe below using symplectic correctors.
}.
It is in general not possible to derive an exact expression for the corrector $e^{ C}$, so it has to be approximated.
We can do this approximation to any order we want and don't need to make it particularly efficient because the correctors are only applied at the beginning and end of the simulation, not during intermediate timesteps.

\cite{Wisdom2006} gives explicit expressions for $e^{C}$ to various order $dt^{p}$ in terms of the already implemented Keplerian motion and kick operators, $e^{ A}$ and $e^{ B}$. 
After applying the correctors, the largest error terms are  $\bo[\epsilon dt^{p+1}]$ and $\bo[\epsilon^2 dt^2]$.
We refer to this integrator as WHC$p$, where $p$ is the order of the corrector.
If we apply a high enough order corrector, for example a 17th order one, then $\bo[\epsilon dt^{18}]$ can be ignored and the dominant error term is $\bo[\epsilon^2 dt^2]$ for all reasonable timesteps.

%%%%%%%%%%%%%%%%%%%%%%%%%%%%%%%%%%%%%%%%%%%%%%%%%
\subsubsection{Kernel method}
Using the 17th order correctors of \cite{Wisdom2006}, the leading order term of the WHC method is, for all practical purposes, $\bo[\epsilon^2 dt^2]$.
Naturally, we would like to get rid of this error term as well. 
The approach taken by \cite{Wisdom1996} is to change the kick step of the WH method, $e^{B}$.
If we refer to $e^{K}=e^{\frac12 A}e^{ B}e^{\frac12 A}$ as the kernel of the WH and WHC$p$ methods and change it $e^{\frac12 A}e^{ B'}e^{\frac12 A}$, then with the right choice of $B'$ and one extra corrector step, one can, at least in theory, eliminate all  $\bo[\epsilon^2]$ terms. 
We refer to this ideal integrator as WHCCKI (WH + first corrector + second corrector + ideal kernel). 
If it were possible to implement this kernel (as well as the exact correctors), its leading error term would be~$\bo[\epsilon^3dt^4]$.

Unfortunately, the ideal $B'$ is another infinite series of Poisson brackets and we cannot (at least not for the $N$-body problem) hope to solve the evolution of the system under it exactly.
Thus, we are once again required to rely on an approximation.
However, because $B'$ itself is a power series in $dt$, we only need to match $B'$ to a finite power in $dt$ to get rid of the leading order $\bo[\epsilon^2 dt^2]$ term.
Note that by doing so  we still keep higher order terms involving $\epsilon^2$, including the now leading term $\bo[\epsilon^2 dt^4]$.
\cite{Wisdom1996} describes three different methods for a practical implementation of such an approximation. 

The first method involves a composition of $e^{A }$ and $e^{B }$ operators.  
One example of a kernel that has the right properties to eliminate the leading order term is 
\begin{eqnarray}
  e^K = e^{\frac58 A}
e^{-\frac16 B}
e^{-\frac14 A}
e^{\frac16 B}
e^{\frac18 A}
e^{B}
e^{-\frac18 A}
e^{-\frac16 B}
e^{\frac14 A}
e^{\frac16 B}
e^{\frac38 A}. \nonumber
\end{eqnarray}
Note that the implementation of this kernel is straightforward if the we already have the $e^{A }$ and $e^{B }$ operators implemented.
However, in contrast to the symplectic correctors, the kernel operator needs to be applied at every timestep and its efficiency is therefore important. 
The force evaluation is typically the slowest part of an integrator because it scales as $\bo[N^2]$ for $N$ particles, whereas all other parts scale as $\bo[N]$. 
The above kernel contains five $e^{B }$ operators and thus five force evaluations per timestep. 
We can therefore expect this method to be approximately five times slower than the WH and WHC$p$ methods.
We will see below if the increase in accuracy outweighs the decrease of performance. 
We will refer to this method as WHCCKC (WH + first corrector + second corrector + composition kernel). 
Because the composition kernel is only accurate to order $\bo[\epsilon^2 dt^4]$, we might ignore the second corrector which only acts at higher orders.
We will refer to this method with only the first corrector applied as WHCKC (WH + first corrector + composition kernel). 

The second method to approximate $B'$ is to calculate the leading order terms explicitly.
For the $N$-body problem, the leading order term, $B$, corresponds to the standard kick, involving derivatives of the planet-planet potentials and some additional terms due to Jacobi coordinates \citep{ReinTamayo2015}. 
The second term in $B'$ is a Poisson bracket of the form $\{B,\{B,A\}\}$.
For a WH type splitting of the $N$-body Hamiltonian in Jacobi coordinates, $B$ is only dependent on positions, and $\{B,\{B,A\}\}$ evaluates to a combination of spatial derivatives of the accelerations (thus second derivatives of the potential) and leads to expressions similar in structure to those describing the jerk\footnote{The jerk is the time derivative of the acceleration.}.
Thus, it is possible to implement an efficient operator corresponding to the evolution under the Hamiltonian
\begin{eqnarray}
    \widehat B = B + dt^2 \frac1{24} \{B,\{B,A\} \}.
\end{eqnarray}
In fact, it is possible to calculate $e^{\widehat B}$ (the `modified kick') without any additional square root evaluations beside those already needed to calculate $e^{B}$ \citep{Wisdom2018}.
Thus, if we use the kernel 
\begin{eqnarray}
  e^K = e^{\frac12 A}
e^{\widehat B}
e^{\frac12 A}
\end{eqnarray}
we can expect a performance almost identical to that of the standard WH method. 
We refer to method which uses this approximation of the kernel as WHCKM (WH + first corrector + modified kick kernel).
If we decide to use the second correctors as well, the method becomes WHCCKM.
As above, the use of second correctors is not expected to improve the accuracy because we already introduce error terms at order $\bo[\epsilon^2 dt^4]$ by truncating the in principle infinite series of $B'$ to only two terms to get $\widehat B$.

The third method to approximate $B'$ is referred to as the `lazy implementer's method' by \cite{Wisdom1996}. 
Instead of calculating the Poisson bracket $\{B,\{B,A\}\}$ explicitly, it uses a Taylor series approximation to estimate the modified kick by first calculating the acceleration $a_j$ due to $e^{B}$ (i.e. the unmodified kick), then shifting the positions with $q_j\rightarrow q_j+\frac{dt^2}{12}a_j$, before calculating new accelerations $a'_j$ with the updated positions, and finally updating the momenta with $p_j\rightarrow p_j+dt\, m_j\, a'_j$.
The intermediate position values are discarded. 
This trick requires two force evaluations for the kernel and the method is thus approximately two times slower than the standard WH method. 
Note that the Taylor series introduces a new error term at order $\bo[\epsilon^3 dt^3]$.
Depending on the relative size of $\epsilon$ compared to $dt$, this may or may not be the new dominant term.
In either case, because the kick step is only updating the momenta, the scheme itself remains symplectic.
Note that compared to the modified kick method, the lazy implementer's method requires no extra work when it is used with any additional position dependent potential, i.e. due to general relativity\footnote{General relativistic corrections are in fact velocity dependent. However, the perihelion precession can be modelled with a velocity independent $r^{-3}$ potential.} or tides.
We refer to this method as WHCKL (WH + first corrector + lazy implementer's kernel). 
Although the use of the second corrector might not be of any advantage here either, for completeness, we refer to it as WHCCKL.

%%%%%%%%%%%%%%%%%%%%%%%%%%%%%%%%%%%%%%%%%%%%%%%%%
\subsection{The SABA integrator family}
\label{sec:SABA}
A different approach to extend the WH integrator to higher order has been used by \cite{LaskarRobutel2001}.
To construct their SABA family of symplectic integrators, these authors use a Lie series approach to build composition methods similar to \cite{Yoshida1993}.
An additional constraint which \cite{LaskarRobutel2001} enforce is that each sub-step in the composition method has only positive timesteps. 
They argue that this avoids large coefficients in the error terms, which might be beneficial for large timesteps. 

Specifically, their integrators SABA1, SABA2, SABA3, and SABA4, have error terms 
$\bo(\epsilon dt^2)$,
$\bo(\epsilon dt^4+\epsilon^2 dt^2)$,
$\bo(\epsilon dt^6+\epsilon^2 dt^2)$, and
$\bo(\epsilon dt^6+\epsilon^2 dt^2)$ respectively.
They are compositions of the same $e^{A}$ and  $e^{B}$ operators used in the classical WH method.
For example, one SABA2 step can be written as
\begin{eqnarray}
    \text{SABA2} = e^{c_1 A}
    e^{d_1 B}
    e^{d_2 A}
    e^{d_1 B}
    e^{c_1 A}
\end{eqnarray}
where the coefficients $c_1$, $d_1$, and $d_2$ are positive constants. 
It is referred to as SABA2 because it involves two force evaluations per timestep. 
Similarly SABA3 has three force evaluations per timestep, and so forth. 
SABA1 is simply the classical WH method which uses one force evaluation.

As \cite{LaskarRobutel2001} point out and we will show later, the dominant term for small timesteps of the SABA methods is typically the $\bo(\epsilon^2dt^2)$ term except for SABA1 (which is just WH). 
To remove this error term, \cite{LaskarRobutel2001} describe a corrector step for their integrators, leaving only terms of order $\bo(\epsilon dt^p + \epsilon^2 dt^4)$ if used in conjunction with the SABA$p$ integrators.
We refer to these integrators as SABAC$p$.

Is it important to stress that the SABAC$p$ correctors are very different from the ones described by \cite{Wisdom1996}, both in terms of concept and implementation.
For example, consider SABAC2 which we can write as
\begin{eqnarray}
    \text{SABAC2} =
    e^{-\alpha C_S}
    e^{c_1 A}
    e^{d_1 B}
    e^{d_2 A}
    e^{d_1 B}
    e^{c_1 A}
    e^{-\alpha C_S},
\end{eqnarray}
where $\alpha$ is a positive constant.
The SABAC$p$ correctors $e^{C_S}$ are simply given by the evolution of the system under the dominant term in the shadow Hamiltonian, $\{B,\{B,A\}\}$ and the coefficient $\alpha$ is directly related to the coefficient of this term as it appears in the shadow Hamiltonian. 
Thus the $e^{\alpha C_s}$ operators explicitly remove one of the error terms accumulated over one timestep.
Note that this is the same Poisson bracket we encountered in the kernel method WHCKM above. 
However, the interpretation is different.
Further note that in contrast to the WHC$p$ correctors which are only applied at the beginning and end of the integration, the SABAC$p$ correctors do not cancel out and have to be applied at every timestep.

To calculate these correctors, we need an operator corresponding to the evolution under the Hamiltonian $\{B,\{B,A\}\}$. 
In the case of the standard $N$-body problem with Newtonian gravity, the operator can be calculated analytically.
In fact, it is the same calculation as for the modified kick in the WHCKM integrator except that we throw away the standard part of the kick and only keep the modifications. 
Thus, this corrector implementation takes about as much time as a normal force evaluation.
This is the approach taken by \cite{LaskarRobutel2001}.

However, since we have encountered different ways to calculate the kernel in the last section, it might come at no surprise that there is another option to calculate the $e^{C_S}$ correctors.
To our knowledge this method has not been described elsewhere. 
It uses the same idea as the lazy implementer's kernel in WHCKL. 
However, rather than advancing the momenta by the modified kick, $p_j\rightarrow p_j+dt\, m_j\, a'_j$, we only advance them by the modification, i.e. $p_j\rightarrow p_j+\beta\, dt\, m_j\, (a'_j-a_j)$ with some constant $\beta(\alpha)$.
As for WHCKL, this method has the disadvantage that it only approximates the evolution under $\{B,\{B,A\}\}$ and will therefore introduce errors at higher order.
But it also has the same advantage in that it works with any arbitrary position dependent force without the need to derive an analytic expression for  $\{B,\{B,A\}\}$.
We refer to this integrator as SABACL (SABA integrator with lazy correctors).

\cite{LaskarRobutel2001} also describe higher order SABA methods, the analogue SBAB integrators where the $e^{A}$ and $e^{B}$ operators are exchanged, as well as even higher order methods that can be constructed by compositions of lower order SABA and SBAB integrators. 
Since these integrators seem to be less efficient and useful for typical $N$-body simulations of planetary systems \citep[the long and highly accurate integrations of][use the SABAC4 method]{La2010}, we do not consider them in this paper.
%%%%%%%%%%%%%%%%%%%%%%%%%%%%%%%%%%%%%%%%%%%%%%%%%
\subsection{SABA integrators with negative timesteps}
\label{sec:SABAneg}
The dominant error term in the SABA4 integrator with correctors is of order $\bo[\epsilon^2 dt^4]$ for small timesteps.
\cite{LaskarRobutel2001} enforced that all sub-steps have positive timesteps.
To achieve a higher order in $dt$, one needs to relax this condition.
Doing so, \cite{Blanes2013} present methods with leading error terms of $\bo[\epsilon dt^{10} + \epsilon^2 dt^4]$, $\bo[\epsilon dt^8 + \epsilon^2 dt^6 + \epsilon^3 dt^4]$, and $\bo[\epsilon dt^{10} + \epsilon^2 dt^6 + \epsilon^3 dt^4]$.
The specific methods chosen by \cite{Blanes2013} have particularly small constants associated with their error terms. 
We refer to these methods as SABA(10,4), SABA(8,6,4), and SABA(10,6,4).
They need 7, 7 and 8 force evaluations per timestep, respectively, but do not require a corrector step.
Therefore these methods work with any position dependent forces.
In contrast to the SABA methods in the last section, The negative timesteps make the method here harder to use with dissipative forces.
\cite{Blanes2013} also discuss methods using heliocentric coordinates.
Since these methods have almost identical efficiency compared to those using Jacobi coordinates, we do not further consider them here.

%%%%%%%%%%%%%%%%%%%%%%%%%%%%%%%%%%%%%%%%%%%%%%%%%
\subsection{Comparison table for integrators}
\begin{table*}
    \caption{
        Comparison of higher order symplectic splitting integrator used in planetary dynamics. 
        Integrators with particularly useful properties have been highlighted with $\bigstar$.
        See text for details. 
     \label{tab:integrators}}
    \begin{center}
        \begin{tabular}{ >{\raggedright}p{2.5cm} >{\raggedright}p{1.3cm} >{\raggedright}p{1.5cm}  >{\raggedright}p{2.2cm}   l p{5mm} >{\raggedright}p{8mm}  >{\raggedright}p{10mm}   >{\raggedright}p{10mm}  p{21mm}  } 
            Name and synonyms& Main references & Start up/ Shut down & One timestep  & Cost & Only $A, B$ & $\bo[\epsilon\, dt^?]$& $\bo[\epsilon^2\, dt^?]$& $\bo[\epsilon^3\, dt^?]$ & Implemented in \reb \\\toprule
            %%%%%%%%%
            WH  $\bigstar$\newline SABA1 (d) \newline WHFAST (e) \newline M2 (b)& (a), (e), (f) & - & $A\,B\,A$ & 1 & 
            \checkmark & 
            2 &  &  
            & \checkmark  \\\midrule 
            %%%%%%%%%
            WHC$p$\newline CM2 (b)& (b), (c), (f) & $C^{(1)}_{[p]}$ & $A\,B\,A$ &  1 & 
            \checkmark & 
            p+1 & 2 & 
            & \checkmark (up to $p=17$)\\\midrule 
            %%%%%%%%%
            WHCCKI \newline (ideal kernel) & (b) & ${C}^{(2)}_*\, {C}^{(1)}_*$& $A\,B_*\,A $ \newline $*=$ not possible & & 
            & 
            $\infty$ & $\infty$ & 4
            &  (not possible)\\\midrule 
            %%%%%%%%%
            WHCKC \newline (comp. kernel) & (b) & $C^{(1)}_{[17]}$ & $A\,(B\,A)^5 $ & 5 & 
            \checkmark & 
            18 & 4&  
            & \checkmark \\\midrule 
            %%%%%%%%%
            WHCKM \newline (mod. kick kernel)\newline CMM4  (b) & (b) & $C^{(1)}_{[17]}$ & $A\,\widehat B\,A $ & 1 & 
            & 
            18 & 4 & 
            &  \checkmark \\\midrule 
            %%%%%%%%%
            WHCKL $\bigstar$\newline (lazy impl.  kernel) &(b) & $C^{(1)}_{[17]}$ & $A\,\widehat{B\,B}\,A $ & 2 & 
            \checkmark & 
            18 & 4 & 3 
            & \checkmark \\\midrule 
            %%%%%%%%%
            WHCCKC\newline (comp. kernel) & (b) & $ C^{(2)}\, C^{(1)}_{[17]}$ & $A\,(B\,A)^5 $ & 5 & 
            \checkmark & 
            18 & 4 & 
            & \checkmark \\\midrule 
            %%%%%%%%%
            WHCCKM \newline (mod. kick kernel) \newline WHCK (f) & (b) (f) & $C^{(2)}\, C^{(1)}_{[17]}$ & $A\,\widehat B\,A $ & 1 & 
            & 
            18 & 4 & 
            & \checkmark  \\\midrule 
            %%%%%%%%%
            WHCCKL \newline(lazy impl.  kernel) &(b) & $ C^{(2)}\, C^{(1)}_{[17]}$ & $A\,\widehat{B\,B}\,A $ & 2 & 
            \checkmark & 
            18 & 4 & 3
            & \checkmark \\\midrule 
            %%%%%%%%%
            SABA2 & (d) & - & $A\,B\,A\,B\,A$ &  2 & 
            \checkmark & 
            4 & 2 &  
            & \checkmark \\\midrule 
            %%%%%%%%%
            SABA3 & (d) & - & $A\,(B\,A)^3$ & 3 & 
            \checkmark & 
            6 & 2 & 
            & \checkmark \\\midrule 
            %%%%%%%%%
            SABA4 & (d) & - & $A\,(B\,A)^4$ &  4 & 
            \checkmark & 
            6 & 2 & 
            & \checkmark \\\midrule 
            %%%%%%%%%
            SABAC2 & (d) & - & $\widehat B A\,(B\,A)^2 \widehat B $ & 3 & 
            & 
            4 & 4 & 
            & \checkmark  \\\midrule 
            %%%%%%%%%
            SABAC3 & (d) & - & $\widehat B A\,(B\,A)^3 \widehat B $ & 4 & 
            & 
            6 & 4 &  
            & \checkmark  \\\midrule 
            %%%%%%%%%
            SABAC4 & (d) (g) & - & $\widehat B A\,(B\,A)^4 \widehat B $ & 5 & 
            & 
            6 & 4 &  
            & \checkmark  \\\midrule 
            %%%%%%%%%
            SABACL4 $\bigstar$\newline(lazy impl. cor.) & this paper & - & $\widehat{B\,B} A\,(B\,A)^4 \widehat{B\,B} $ & 6 & 
            & 
            6 & 4 & 3
            & \checkmark \\\midrule 
            %%%%%%%%%
            SABA(10,4)  & (h) & - & $A\,(B\,A)^7  $ & 7 & \checkmark 
            & 
            10 & 4 & 
            & \checkmark \\\midrule 
            %%%%%%%%%
            SABA(8,6,4)  & (h) & - & $A\,(B\,A)^7  $ & 7 & \checkmark 
            & 
            8 & 6 & 4 
            & \checkmark \\\midrule 
            %%%%%%%%%
            SABA(10,6,4)   $\bigstar$ & (h) & - & $A\,(B\,A)^8  $ & 8 & \checkmark 
            & 
            10 & 6 & 4
            & \checkmark \\\midrule 
        \end{tabular}\\
        (a) \citealt{WisdomHolman1991}, (b) \citealt{Wisdom1996}, (c) \citealt{Wisdom2006}, (d) \citealt{LaskarRobutel2001}, \\(e) \citealt{ReinTamayo2015}, (f) \citealt{Wisdom2018}, (g) \citealt{La2010}, (h) \citealt{Blanes2013}
    \end{center}
\end{table*}

Whereas the WH method is well known by astronomers and the go-to method for long term integrations of planetary systems, the higher order methods are used less often.
The main motivation behind writing this paper is to clear up some of the mystification regarding the various higher order integrators and make them readily available for astronomers to use.

To do that we summarize all integrators that we have introduced above in Table.~\ref{tab:integrators}. 
The first column lists the name of the methods and synonyms various authors have used for them.
The second column lists the main reference(s) for each integrator. 
The third column lists the operators that need to be applied to start up and shut down an integrator.
$C^{(1)}_{p}$ indicates first symplectic correctors of order $p$ and $C^{(2)}$ indicates second symplectic correctors.
The fourth column lists the operators which need to be applied at each time step. 
Note that in some cases it is possible to combine the last operator of the previous step with the first operator of the current step.
The symbol $A$ corresponds to $e^{A}$, the evolution under Hamiltonian $A$ which describes the Keplerian motion of planets.
Similarly, $B$ corresponds to $e^{B}$, the planet-planet interactions or the kick step.
For integrators which use a variant of the modified kick operator, we denote $\widehat B$.
Similarly, for integrators which use a variant of the lazy implementer's method, we denote $\widehat{BB}$ (two $B$'s because it involves two force evaluations).

We list the theoretical cost of each method in column five. 
It corresponds to the number of force evaluations per timestep and assumes all other operations take no time.
If the last operator of the step involves a force evaluation and it can be combined with the first operator of the next step, then we only count it as one force evaluation.
This column thus provides a runtime estimate relative to the classical WH method for a fixed timestep. 
We come back to this later, but note that two methods already stands out: WHCKM and WHCCKM. 
We should expect these integrators to be as fast as the classical WH method.

The sixth column lists if the method requires only the operators $e^{A}$ and $e^{B}$.
If this is the case, then the implementation is straightforward and just comes down to repeatedly applying  $e^{A}$ and $e^{B}$ in the right order and for the right amount of time. 
Note that if other operators need to be implemented, then it is harder to use the method for simulations which include effects other than Newtonian gravity between all particles because these special operators need to be rederived analytically\footnote{An auto-differentiation algorithm could do this too.}.

The seventh, eighth, and ninth columns list the order of the leading error terms.
A number $k$ in the seventh column implies that there is an error term of order $\bo[\epsilon dt^k]$, 
a number $k$ in the eighth column implies that there is an error term of order $\bo[\epsilon^2 dt^k]$, and similarly for the ninth column. 
A column is left blank if the error is always dominated by the error given by the column to the left, regardless of the relative sizes of $dt$ and $\epsilon$.

The last column lists if the method is implemented in the \reb integrator package.
We will describe the details of our implementations in the next section.

%%%%%%%%%%%%%%%%%%%%%%%%%%%%%%%%%%%%%%%%%%%%%%%%%
%%%%%%%%%%%%%%%%%%%%%%%%%%%%%%%%%%%%%%%%%%%%%%%%%
\section{Implementation}
\label{sec:implementation}

%%%%%%%%%%%%%%%%%%%%%%%%%%%%%%%%%%%%%%%%%%%%%%%%%
\subsection{WHFast extensions}
We implement all methods described in Sec.~\ref{sec:WHT} as extensions to the \whfast integrator in \reb.
To do this we add two new parameters to the \texttt{ri\_whfast} structure. 
The first is \texttt{kernel} which controls what kind of kernel is used.
The default setting (\texttt{DEFAULT}) uses the standard WH method's kernel. 
The other settings available are  \texttt{COMPOSITION}, \texttt{MODIFIEDKICK}, and \texttt{LAZY}, corresponding to the integrators WHCKC, WHCKM, and WHCKL, respectively.
The second new parameter is \texttt{corrector2} which either turns the second correctors on or off.
We also extend the first correctors already implemented in \whfast and controlled by the \texttt{corrector} parameter to 17th order.

The new kernel methods are currently implemented in a way that makes them somewhat more restricted in their usage than the basic \whfast algorithm. 
Specifically, the newly implemented kernels do not support variational equations (and therefore chaos indicators such as MEGNO), OpenMP parallelization, or force calculations using a BH tree code. 
For most cases where higher order symplectic methods might be used, these features are not essential.
However, some of these features can be added at a later time if there is a need.

All kernel methods make use of the \whfast setting \texttt{safe\_mode} which, when turned off, combines the last $e^A$ operator in each timestep with the $e^A$ operator at the beginning of the next timestep.
If the safe mode is turned off, then Jacobi coordinates are only converted back to cartesian coordinates at the end of the integration.
In most cases it makes sense to turn off the safe mode flag as it not only provides a speed-up, but also reduces round-off errors coming from continuous transformations to and from Jacobi coordinates.

If frequent outputs are required \whfast needs to apply the correctors and their inverses repeatedly. 
Round-off error can prevent the correctors and their inverses from cancelling out exactly.
If this becomes a problem, one can turn on the \texttt{keep\_unsynchronized} setting of \whfast which then only applies the correctors to generate an output in cartesian coordinates, but continues the integration from a copy of the Jacobi coordinates it made before the correctors were applied.

\reb allows the user to specify a routine to include additional forces which can be used to model effects due to general relativistic corrections, oblateness, or tides.
As long as these forces are position dependent, all of the new kernel methods support additional forces, with the exception of \texttt{MODIFIEDKICK}. 

Furthermore, all new kernels are compatible with the SimulationArchive \citep{ReinTamayo2017}.
Specifically, all kernel methods are bit-wise reproducible from any snapshot and are machine independent.

%%%%%%%%%%%%%%%%%%%%%%%%%%%%%%%%%%%%%%%%%%%%%%%%%
\subsection{SABA}
We implement the SABA integrator family described in Sections~\ref{sec:SABA} and~\ref{sec:SABAneg} as a new integrator \texttt{saba} in \reb.
The parameter \texttt{type} in the \texttt{ri\_saba} structure determines which specific SABA integrator is used.
Currently, the three high order integrators of \cite{Blanes2013}, SABA(10,4), SABA(8,6,4), and SABA(10,6,4) are implemented.    
In addition the integrators SABA1, SABA2, SABA3, and SABA4 with either no correctors, lazy correctors, or modified kick correctors are implemented.
It is straightforward to extend the implemention to other variants should there be a need.

Both the \texttt{keep\_unsynchronized} and the \texttt{safe\_mode} parameters in the \texttt{ri\_saba} structure works the same way as for \whfast.
If the safe mode is turned off, it combines the $e^A$ operators at the beginning and end of consecutive timesteps if no corrector is used.
If a corrector is used, then the correctors at the beginning and end of a timestep are combined.
As with \whfast, in most cases it makes sense to turn off the safe mode flag to provide a speed-up and reduce round-off errors.
Note that it is called safe mode because, when turned off, changing particle properties manually in-between timesteps requires extra scrutiny. 

The modified kick corrector is not compatible with additional forces, however, SABACL integrators using the lazy correctors are, as are all the SABA integrator which do not use a corrector.

The SABA integrators as implemented in \reb rely on many of the internal \whfast methods and are therefore also compatible with the SimulationArchive, bit-wise reproducible, and machine independent.

%%%%%%%%%%%%%%%%%%%%%%%%%%%%%%%%%%%%%%%%%%%%%%%%%
%%%%%%%%%%%%%%%%%%%%%%%%%%%%%%%%%%%%%%%%%%%%%%%%%
\section{Tests}
\label{sec:tests}

\begin{figure*}
    \centering
    \resizebox{0.99\textwidth}{!}{\includegraphics[trim=0cm 0cm 0cm 0cm]{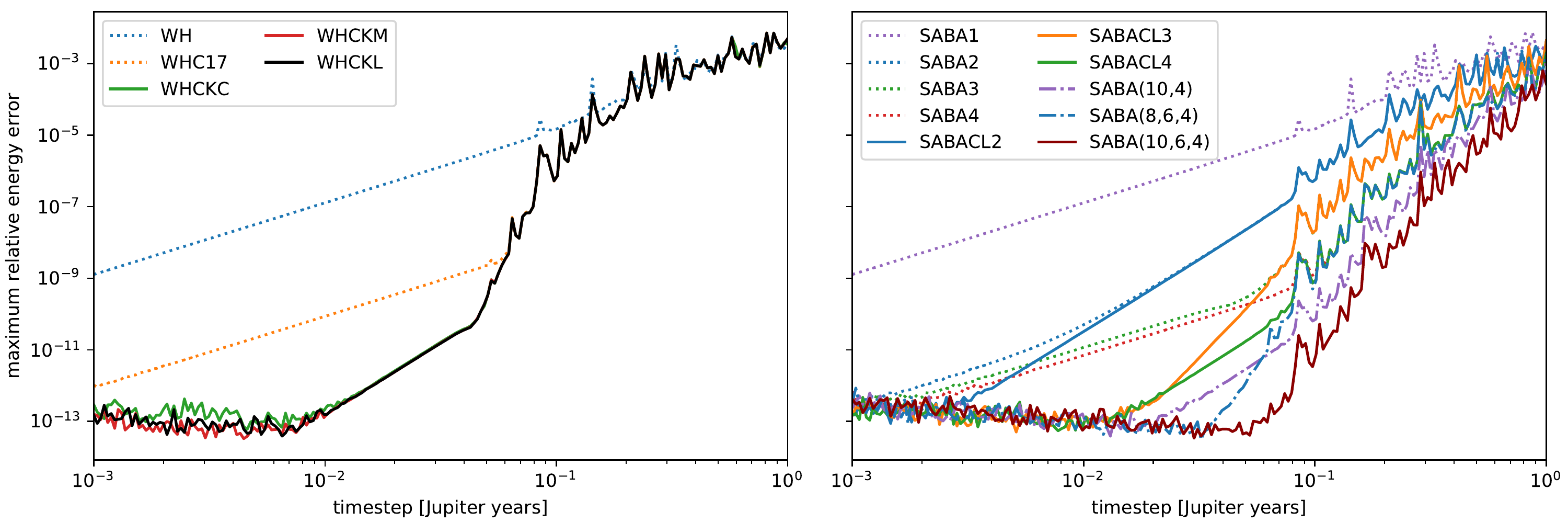}}
    \caption{The maximum relative energy error as a function of the timestep in 10~kyr integrations of the outer Solar System using different symplectic integrators. 
    \label{fig:energy}
    }
\end{figure*}

\begin{figure*}
    \centering
    \resizebox{0.99\textwidth}{!}{\includegraphics[trim=0cm 0cm 0cm 0cm]{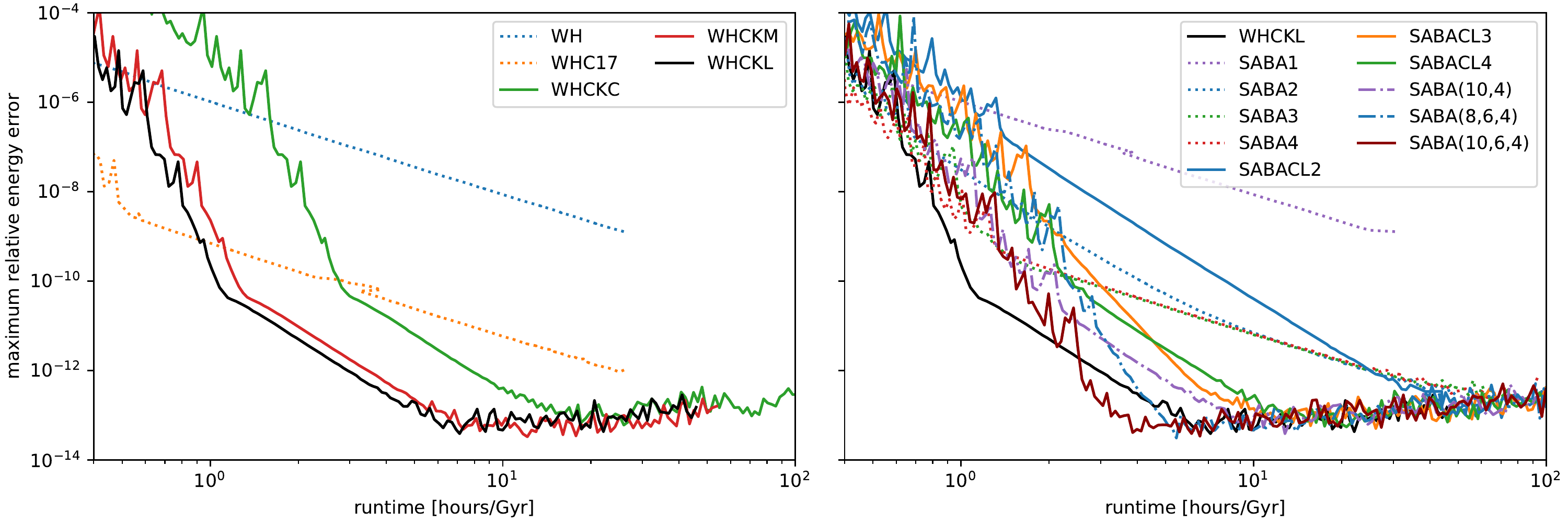}}
    \caption{The maximum relative energy error in simulations of the outer Solar System as a function of the runtime required to reach~1Gyr using different symplectic integrators. 
    \label{fig:speed}
    }
\end{figure*}

We run simulations of the outer Solar System for 10kyrs as a test case of our \reb implementations of the integrators introduced above.
A python notebook to reproduce all of the test and figures can be found at \url{https://github.com/hannorein/ReinTamayoBrown2019}.

Fig.~\ref{fig:energy} shows the maximum relative energy error as a function of the timestep.
We measure the energy error at 10,000~random times during each integration to avoid any aliasing. 
The left panel shows the methods of \cite{Wisdom1996} whereas the right panel shows those of \cite{LaskarRobutel2001} and \cite{Blanes2013}.
All methods are dominated by stepsize resonances \citep{Rauch1999} for timesteps larger than about 10\% of the shortest orbital timescale in the problem. 
Note that for such large timesteps, the SABA type integrators perform better than others. 
The first reason for this is that the SABA$p$ methods with $p>1$ have more than one force evaluation during the timestep.
Thus, the effective timestep in-between force evaluations is smaller than the timestep plotted on the horizontal axis.
Of course, more force evaluations comes at the cost of being slower (see below). 
The second reason for the good behaviour of the SABA integrators at large timesteps is that the coefficients in the error terms are generally small even at higher order which is beneficial for large timesteps \citep{LaskarRobutel2001}. 

For more reasonable timesteps of a few percent of the shortest orbital period, the methods show a power-law convergence as indicated by the terms in Table~\ref{tab:integrators}.
In particular, note that the WH and WHC17 integrators follow a $\bo[\epsilon dt^2]$ and $\bo[\epsilon^2 dt^2]$ power-law, respectively.
In other words, the error of WHC17 is reduced by one extra factor of $\epsilon$ compared to the standard WH integrator.
From the plot, we can read off a value $\epsilon\sim10^{-3}$ for this test case, roughly equal to the Jupiter-Sun mass ratio.
For small enough timesteps, once the $\bo[\epsilon]$ term is negligible, the SABA2, SABA3, and SABA4 integrators\footnote{Except SABA1 which is equivalent to the standard WH method.} follow the same $\bo[\epsilon^2 dt^2]$ power law as WHC17.
In other words, going beyond SABA2 does not help to improve the accuracy significantly for small timesteps.
Note however that SABA3 is slightly more accurate than SABA2, and SABA4 is slightly more accurate than SABA3. 
Once again, this can be explained because SABA3 evaluates the forces three times during a timestep, whereas SABA2 evaluates them only twice. 
Thus, the effective timestep of SABA3 is actually smaller than that of SABA2.
The correctors of the SABAC integrators successfully remove the $\bo[\epsilon^2 dt^2]$ terms, with higher order SABAC integrators having an advantage at large timesteps since the leading order errors fall off faster for higher-order SABA integrators.
The higher order integrators of \cite{Blanes2013} have even smaller energy errors.

Going back to the left panel of Fig.~\ref{fig:energy}, we can see that all kernel methods, WHCKC, WHCKM, and WHCKL, perform equally well. 
Thus, we conclude that any of the three approximations for the kernel work equally well for this problem, which is consistent with the results of \cite{Wisdom1996}. 

For very small timesteps, the integrators are dominated by numerical round-off error coming from the finite precision of floating point numbers. 
If all round-off errors are unbiased, then this error term behaves like a random walk and scales as $\bo(dt^{-1/2})$.
This is known as Brouwer's law \citep{Brouwer1937}.
This behaviour can be observed for all integrators in Fig.~\ref{fig:energy} that reach machine precision and is expected as all integrators use internally the unbiased \whfast Kepler solver \citep{ReinTamayo2015}.
Worth noting is that the WHCKC method has a slightly larger round-off error than the WHCKM and WHCKL methods because it requires more $e^A$ and $e^B$ operators for one timestep.

To better compare the runtime performance of the integrators, we plot the maximum relative energy error as a function of the runtime in Fig.~\ref{fig:speed}.
The simulations were performed on an Intel Xeon CPU (E5-2620 v3, 2.40GHz). 
The horizontal axis has been scaled so that it shows the number of hours required for a 1 billion year integration for this particular problem and CPU.
Amongst all integrators, WHCKL is the fastest for moderate accuracy, $\Delta E/E\gtrsim 10^{-12}$, (we plot it on both panels for comparison).
It only requires one hour to integrate the outer Solar System for 1 Gyr at a maximum relative energy error of $10^{-10}$.
Since the dominant error term, $\bo[\epsilon^2 dt^4]$, falls off as the fourth power of the timestep, an accuracy of $10^{-14}$ can be achieved in only 10~hours. 
The WHCKM integrator with the modified kick step performs almost as well.
For simulations which require extremely high accuracy, the higher order SABA integrators perform best. 
In particular, the SABA(10,6,4) integrator is more efficient than the WHCKL method when $\Delta E/E\lesssim 10^{-12}$ is required. 
It requires a runtime of four hours to integrate the outer Solar System for 1 Gyr at a maximum relative energy error of $10^{-14}$, roughly a factor of 2 faster than WHCKL. 

The fastest integrators of the SABA family for high accuracy simulations is SABA(10,6,4).
We can see that for the same accuracy, the fastest SABACL/SABAC integrator is roughly a factor of 2-5 slower than the fastest integrators in our sample, WHCKL and SABA(10,6,4).
This is consistent with the results of \cite{Wisdom2018}.

Not shown in the plots are the integrators of \cite{Wisdom1996} which use second symplectic correctors, WHCCKM, WHCCKL, and WHCCKC.
We do not observe any improvement of the energy error in this test case compared to their counterparts with only first symplectic correctors applied.
We also do not show the SABAC integrators because they perform similarly to their SABACL counterparts.

%%%%%%%%%%%%%%%%%%%%%%%%%%%%%%%%%%%%%%%%%%%%%%%%%
%%%%%%%%%%%%%%%%%%%%%%%%%%%%%%%%%%%%%%%%%%%%%%%%%
\section{Conclusions}
\label{sec:conclusions}

In this paper, we reviewed different high order symplectic integrators for long term direct $N$-body simulations of planetary systems. 
We implemented all of these integrators and make them freely available as part of the \reb integrator package.
  Some of these method have a truly remarkable performance with little to no additional cost associated when compared to the (already impressive) standard second order Wisdom-Holman method.
For example, a typical integration of the Solar System which takes the WHCKL method 10~hours to complete, would require more than a year if one were to use the standard Wisdom-Holman method and require the same level of accuracy.

The best performing integrators in our sample, WHCKL, SABA4CL, and SABA(10,6,4) use very different approaches to achieve a high accuracy. 
Reassuringly, the different approaches lead to integrators with similar performance. 
We find a that the WHCKL integrator has a small\footnote{`Small' depends on the context. If one has to wait for a year for a simulation to finish, a factor three increase in performance might be a huge deal.} advantage over the best integrators in the SABA family for moderate accuracies, being about 2-3 times faster. 
On the other hand, the SABA(10,6,4) integrator is faster for very high accuracy runs, $\Delta E/E\lesssim 10^{-12}$.

Aside of speed and accuracy, the integrators differentiate in other ways as well. 
The SABA, SABACL, WHCKC, and, WHCKL integrators only require operators that are already present in the standard WH integrator, i.e. a Kepler solver and a routine calculating the interaction terms. 
This makes their implementation very straightforward.
Furthermore, they can be used together with forces other than Newtonian gravity, as long as these forces only depend on the particles' positions. 
Astrophysically relevant forces with such a property include general relativistic corrections, quadrupole and other higher order moments of non-spherical objects, and some descriptions of tidal and radiation effects.
The other integrators, SABAC and WHCKM, can also be used for these cases, but some work is needed in addition to the force implementation itself. 

Our implementations do not support extended precision.
For very small timesteps, finite precision of floating point numbers is therefore the limit factor in achieving even higher precision.
There are currently few problems where this level of precision is required.
One exception are Solar System integrations where our understanding of the physical system is now comparable to this level of precision \citep{La2010}. 
Either a calculation in full quadruple precision or some form of compensated summation \citep{Wisdom2018} can be used to go beyond the limits of double precision floating point numbers.

In summary, for integrations of planetary systems in which orbits remain well separated, we recommend WHCKL, the Wisdom-Holman method with the lazy implementer's kernel and first symplectic correctors of order~17. 
To use WHCKL in \reb, one can simply set \texttt{sim.integrator = "WHCKL"}, which configures the \whfast parameter such that it corresponds to WHCKL\footnote{The settings \texttt{safe\_mode} and \texttt{keep\_unsynchronized} are not changed by this shortcut.}. 
The speed of this method is very similar to the standard WH method, but the accuracy is superior in almost all cases and it can be used with a wide variety of additional forces \citep{Tamayo2019}. 
For very high accuracy runs, we recommend the SABA(10,6,4) integrator.
To use SABA(10,6,4) in \reb, simply set \texttt{sim.integrator = "SABA(10,6,4)"}.
For systems in which close encounters occur, a different approach is needed \citep{ReinSpiegel2015, Rein2019}.

%%%%%%%%%%%%%%%%%%%%%%%%%%%%%%%%%%%%%%%%%%%%
%%%%%%%%%%%%%%%%%%%%%%%%%%%%%%%%%%%%%%%%%%%%
\section*{Acknowledgments}
This research has been supported by the NSERC Discovery Grant RGPIN-2014-04553 and the Centre for Planetary Sciences at the University of Toronto Scarborough.
Support for this work was provided by NASA through the NASA Hubble Fellowship grant HST-HF2-51423.001-A awarded  by  the  Space  Telescope  Science  Institute,  which  is  operated  by  the  Association  of  Universities  for  Research  in  Astronomy,  Inc.,  for  NASA,  under  contract  NAS5-26555.
This research was made possible by the open-source projects 
\texttt{Jupyter} \citep{jupyter}, \texttt{iPython} \citep{ipython}, 
and \texttt{matplotlib} \citep{matplotlib, matplotlib2}.

\bibliography{full}

\end{document}